\documentclass[a4paper]{article}
\pdfoutput=1
\usepackage[margin=3cm]{geometry}
\usepackage{graphicx}
\usepackage[utf8]{inputenc}
\usepackage{amssymb}
\usepackage{multirow}
\usepackage[numbers]{natbib}
\usepackage{bm}
\usepackage{authblk}

\begin{document}

\title{Socially Driven News Recommendation}

\author[1]{Nuno Moniz\thanks{nmmoniz@inescporto.pt}}
\author[1]{Luís Torgo\thanks{ltorgo@dcc.fc.up.pt}}
\author[2]{Magdalini Eirinaki\thanks{magdalini.eirinaki@sjsu.edu}}
\affil[1]{LIAAD - INESC Tec\\ FCUP - DCC, University of Porto}
\affil[2]{Department of Computer Engineering, San Jose State University}

\renewcommand\Authands{ and }

\date{December 31, 2015}

\maketitle

\begin{abstract}
The participatory Web has enabled the ubiquitous and pervasive access of information, accompanied by an increase of speed and reach in information sharing. Data dissemination services such as news aggregators are expected to provide up-to-date, real-time information to the end users. News aggregators are in essence recommendation systems that filter and rank news stories in order to select the few that will appear on the user’s front screen at any time. One of the main challenges in such systems is to address the recency and latency problems, that is, to identify as soon as possible how important a news story is. In this work we propose an integrated framework that aims at predicting the importance of news items upon their publication with a focus on recent and highly popular news, employing resampling strategies, and at translating the result into concrete news rankings. We perform an extensive experimental evaluation using real-life datasets of the proposed framework as both a stand-alone system and when applied to news recommendations from Google News. Additionally, we propose and evaluate a combinatorial solution to the augmentation of official media recommendations with social information. Results show that the proposed approach complements and enhances the news rankings generated by state-of-the-art systems.

\end{abstract}

\section{Introduction}\label{sec:intro}

In the current era of participatory Web, dissemination of information has become prevalent. The ubiquitous and pervasive access of information was accompanied by the increase of speed and reach in information sharing. This abundance of information has made the need for efficient data exploration more prominent. The creation and proliferation of platforms which enable users to post information onto the web in real-time and interact with others has brought new challenges into light. As real-time micro-blogging and other social network applications become the main channels of quick information dissemination, traditional information retrieval (IR) and web page ranking approaches have presented several shortcomings in terms of identifying the most relevant information for the end user~\cite{DeChoudhury2011}.

Data dissemination services such as news aggregators are expected to provide up-to-date, real-time information to the end users. 
News aggregators are in essence recommendation systems that filter and rank news stories in order to select the few that will appear on the user's front screen at any time. One of the main challenges in such systems is to address the recency problem, in other words, to identify, as soon as possible, how important a news story is in order to rank it appropriately. The signals used by the news recommendation engine differ depending on the system. While the exact implementation details are undisclosed, we have some insights on how the most  news aggregators, like Google News\footnote{http://news.google.com} work. Google News' recommender system employs a combination of signals including the ``authority'' of the source, the ``international diversity'', the language of the news story, as well as it's original publication time~\cite{Curtiss2012}. Such ``traditional'' news aggregators are slower to identify trending news. On the other hand, participatory media such as Digg\footnote{http://digg.com} incorporate information provided by its own users in real-time (\textit{i.e.} positive votes to a given news item), as well as considering the amount of times shared on Facebook and Twitter. Micro-blog platforms such as Twitter\footnote{http://www.twitter.com} and Facebook\footnote{http://www.facebook.com} serve as a dynamic, multi-topic, real-time feed of world-wide opinion. Such platforms can be used as tools that gauge the importance of a news story as perceived by the public opinion. However, the quality of the sources may be questionable. 

Recently, several works have attempted to address these issues of latency and recency, with the most predominant approach being the use of real-time user input (e.g. number of tweets) as an additional signal to the prediction and recommendation process \cite{Phelan2009,Hsieh2013}. However, such systems introduce some waiting period, as they still need to collect a substantial amount of data in order to make accurate predictions. 

In this work we attempt to alleviate this problem by proposing an integrated framework that predicts the importance of a news item upon its publication, without taking into consideration any feedback from the readers, and translates the resulting predictions to news rankings. To achieve this, we leverage our previous work on resampling strategies \cite{Moniz2014}. Using a real-life data set, collected over the period of 8 months on a half-hour basis, we build the predictive models and translate the results into concrete news rankings. We focus on achieving high prediction accuracy for recent and highly popular news, which is the key distinguishing aspect of our work when compared with previous work.

The proposed solution can be deployed as a stand-alone solution, or integrated in a news recommendations' framework that combines the best of both worlds, combining the news recommendations provided by two types of sources: (i) official media; and (ii) the recommendations of Internet users as they emerge from their social network activity. 

In the framework instantiation presented in this article, we use Google News as the official media source and the number of tweets of a given news item as an indication of how news consumers perceive the importance of a given item.

This approach requires that at any point in time we are able to anticipate how important a news will be in Twitter. If some time has already passed since the news publication, this can be estimated by looking at the observed number of times this news item was tweeted. However, when the news is very recent, the number of already observed tweets will potentially under-estimate the attributed importance of the news item. In this context, for very recent news we need to be able to estimate their future relevance within Twitter, \textit{i.e.} their future number of tweets. Moreover, given that we are interested in using this estimated number of tweets as a proxy for the news relevance, we are only interested in accurately forecasting this number for news with high impact. Additionally, we develop a methodology to evaluate the translation of the resulting predictions into concrete rankings.

In this context, the main contribution of this work is an integrated approach to the problem of accurately predicting and recommending highly popular news upon their publication.  We evaluate our proposed framework both as a standalone system (\textit{i.e.} using a news pool of recent news) and also combined with fixed news recommendations from Google News. Our findings show that, while both systems perform well in specific timeframes and conditions (highlighting their inherent differences, as discussed before), the enhancement of traditional news' rankings with social information improves the rankings and thus the recommendations provided to the end user. This is in accordance to previous work that has employed social network data, but, contrary to existing approaches, the proposed methodology does not require any post-publication input to achieve this.

The rest of this paper is structured as follows. In Section \ref{sec:previouswork} previous work concerning recommender systems, which combine official media and social recommendations as well as prediction approaches, is presented. In Section \ref{sec:probdescr} the data mining tasks are presented and the strategy to handle imbalanced distributions is outlined. Section \ref{sec:datamethods} describes the used data, the regression methods used and the evaluation procedure. The results are discussed in Section \ref{sec:expeval} and conclusions are presented in Section \ref{sec:conclusions}. 

\section{Previous Work}\label{sec:previouswork}

Given a very astounding set of alternatives, recommender systems provide the ability to assist the users in suggesting what the best choices should be. As previously noted that the concrete process of rank production of the previously mentioned popular news recommender systems such as Google News and Digg is not public, although there are some clues available. For example, concerning Google News, official sources state that it is based on characteristics such as freshness, location, relevance and diversity. This process, based on the Page Rank algorithm explained by Page et al. \cite{Page1998} and generally described by Curtiss et al. \cite{Curtiss2012}, is of the most importance as it is responsible for presenting the best possible results for a user query on a set of given terms. However some points have been questioned, such as the type of documents that the algorithm gives preference to, and its effects, and some authors conclude that it favours legacy media such as print or broadcast news ``over pure players, aggregators or digital native organizations'' \cite{Filloux2013}. Also, to the best of our knowledge, this algorithm does not make much use of the available information concerning the impact in or importance given by real-time users, in an apparent strategy of deflecting attempts of using its capabilities in one's personal favour. Nonetheless, the large amount of data that is made available via Twitter enables everyone to access the necessary information for profiling and recommendation processes. The following paragraphs depict examples of those possibilities related to our research scope.

Phelan et. al \cite{Phelan2009} describe an approach to news recommendation that includes harnessing real-time information from Twitter in order to promote news stories. For this endeavor, the authors achieve a basis for story recommendation by mining Twitter data, identifying emerging topics of interest and matching them with recent news coverage from RSS feeds. The approach adopts a content-based recommendation technique and it enables three recommendation strategies: public-rank, friends-rank and content-rank. This work is extended in \cite{Phelan2011} through the increase in comprehension and robustness of the recommendation framework, using different sources of recommendation knowledge and strategies.

Abrol and Khan \cite{Abrol2010} proposes TWinner, a tool capable of combining social media to improve quality of web search and predicting whether the user is looking for news or not. This is done through a process of mining Twitter messages in order to add terms to the search query, a query expansion approach, in order to point out to the search engine when the user is looking for news. This process includes the assignment of weights, measuring of semantic similarity and choosing the $k$ optimum keywords.

The referred approaches have similarities to the one proposed in this paper. The main difference is that in every case, the issue of latency is not addressed any further. Nevertheless, the referred approaches present various options for discussion regarding the combination of official media and social recommendations. 

Determining the importance of a given news article is a very interesting variable when referring to news-based recommender systems. This has been pursued by combining documents from legacy media sources such as newspapers, and the produced content in social media by their users. As previously mentioned, a good example of this is the Digg platform, which continuously provides this correlation between the news articles available and their respective attention. This process gathers enough information to ponder the recommended content according to its popularity or attributed importance. Nevertheless, Digg's algorithm does not provide a prediction of the future importance that the documents will have, to the best of our knowledge. 

Concerning this aspect, of predicting the importance of a news story, some work has been carried out.
Yang and Leskovec~\cite{Yang2011} suggest that popular news could take up to 4 days until their popularity stagnates. Our observations based on previously collected data suggest that the growth of the number of tweets of a news item very rarely exceeds a period of two days, and when it occurs, it is residual.

Regression, classification and hybrid approaches are used by Gupta et. al \cite{Gupta2012} to predict event popularity. In this work the authors use data from Twitter, such as the number of followers/followees. The objective is the same in the work of Hsieh et. al \cite{Hsieh2013}, but the authors approach the problem by improving crowd wisdom with the proposal of two strategies: combining crowd wisdom with expert wisdom and reducing the noise by removing ``overly talkative'' users from the analysis.

Recently, a Bayesian approach was proposed by Zaman et al. \cite{Zaman2013} where a probabilistic model for the evolution of retweets was developed. This work differs from others in a significant manner due to its focus on the prediction of the popularity of a given tweet and its republications (retweets). The authors conclude that the number of retweets of a given tweet after two hours of its publication should achieve a fraction (50\%) of the number of retweets in the first 10 minutes. The test cases include both famous and non-famous Twitter accounts. This work is preceded by others also using the retweet function as a predictor, having as the objective result an interval \cite{Hong2011} or the probability of being retweeted. The dynamics of the retweeting function are discussed in the work of Rudat et al.~\cite{Rudat2014}. Our work assumes a stochastic view of the publications as we count all publications of tweets and not only those that were republished (\textit{i.e.} retweeted), focusing on those that are (predicted to be) highly tweeted.

In the work of Bandari et. al \cite{Bandari2012} classification and regression algorithms are examined in order to predict popularity, translated as the number of tweets, of articles in Twitter. The distinguishing factor of this work from others that attempt to predict popularity of events (\cite{Szabo2010,Lee2010,Kim2011,Lerman2010}), is that it attempts to do this prior to the publication of the item. For this purpose, the authors used four features: source of the article, category, subjectivity in the language, and named entities mentioned. Furthermore, the authors conclude that the source of a given article is one of the most significant predictors. 

The interest of determining the future importance of a given real-time event is mainly related to the quality of recommendations made to the end-user. As stated in most of the work done concerning news recommender systems and Twitter, the distribution of news articles in the referred social network and their number of tweets are described by a power-law distribution~\cite{Tatar2014}. This is an important detail as it shows that only a very small portion of the cases are in fact highly tweeted, and therefore deemed important by the public. And although most of the referenced work obtains encouraging results, none is focused on these rare cases.

The current paper extends our previous work~\cite{Moniz2014} by significantly increasing the amount of data analysed, but mainly by proposing a framework for the translation of the predictions into news rankings and its respective evaluation. Moreover, we approach the problem of combining official and social media in order to ascertain the possibility of augmenting the official media recommendations with social information, aiming at providing better solutions for tackling latency issues related to the recency of news items.

\section{Problem Description and Approach}\label{sec:probdescr}

The objective of our work is twofold: a) to accurately predict the importance of news stories upon their publication, focusing on the news that will be highly tweeted, and b) to transform this outcome into news rankings.

The first issue is a numeric prediction task where we are trying to forecast this number based on some description of the news. However, this task has one particularity: we are only interested in prediction accuracy at a small sub-set of the news - the ones that are tweeted the most. These are the news that the public deems as highly relevant for a given topic and these are the ones we will place at the top of our news recommendation. The fact that we are solely interested on being accurate at a low frequency range of the values of the target variable (the number of tweets) creates serious problems to standard numeric prediction methods that are focused on maximising average accuracy. 

The second task consists on the generation and evaluation of news rankings using the results of the first task. This second task should confirm the possibility of reducing the latency issue related to the recency of news.

In this paper we describe and test several approaches aimed at improving the predictive performance on the difficult task of predicting the importance of a news item. We then apply these prediction models in a framework and evaluate them in two scenarios: \textit{i)} as a standalone system where a news pool of recent news is provided, and \textit{ii)} its application to ranks provided by Google News. The objective of the evaluation is to ascertain the ability of our framework in correctly predicting the most popular news at any given time. Additionally, we attempt to combine our framework with Google News rankings in order to improve the ability to tackle latency and recency issues (\textit{i.e.} enable a faster recommendation of highly popular news).

\subsection{Data Mining Tasks}\label{subsec:dataminingtask}

Concerning the first task, our goal of forecasting the number of tweets of a given news is a numeric prediction task, usually known as a regression problem. This means that we assume that there is an unknown function that maps some characteristics of the news into the number of times this news is tweeted, \textit{i.e.} $Y = f(X_1, X_2, \cdots, X_p)$, where $Y$ is the number of tweets in our case, $X_1, X_2, \cdots, X_p$ are features describing the news and $f()$ is the unknown function we want to approximate. In order to obtain an approximation (a model) of this unknown function we use a data set with examples of the function mapping (known as a training set), \textit{i.e.} $D=\{\langle \mathbf{x}_i, y_i\rangle\}_{i=1}^n$. 

The standard regression tasks we have just formalised can be solved using many existing algorithms, and most of them try to find the model that optimises a standard error criterion like the mean squared error. What sets our specific task apart is the fact that we are solely interested in models that are accurate at forecasting the rare and high values of the target variable $Y$, \textit{i.e.} the news that are highly tweeted. Only this small subset of news is relevant for our overall task of providing a ranking of the most important news for a given topic. In fact, predictive accuracy at the majority of news which have a small number of tweets is completely irrelevant, as such news should not be selected to appear on the front page of a news aggregator.

Regarding the second task, rankings are produced using the outcome of the first task, the predicted number of tweets for each given news item in a certain batch of news. This means that, given these predicted numbers of tweets, this second task executes the trivial process of ranking the news by decreasing predicted number of tweets.

\subsection{Handling the Imbalanced Distribution of the Number of Tweets}\label{subsec:resample}

Previous work \cite{Ribeiro2011,Torgo2007,Torgo2013} has shown that standard regression tools fail dramatically on tasks where the goal is accuracy at the rare extreme values of the target variable. One of the goals of the current paper is to compare some of the proposed solutions to this type of imbalanced regression tasks in the particular problem of forecasting the number of tweets of news.

Several methodologies were proposed for addressing this type of tasks. Resampling methods are among the simplest and most effective. Resampling strategies work by changing the distribution of the available training data in order to meet the preference bias of the users. Their main advantage is that they do not require any special algorithms to obtain the models - they work as a pre-processing method that creates a "new" training set upon which one can apply any learning algorithm. In this paper we will experiment with two of the most successful resampling strategies: (i) SMOTE~\cite{CBOK02} and (ii) under-sampling~\cite{KM97}. These methods were originally developed for classification tasks where the target variable is nominal. 

The basic idea of under-sampling is to decrease the number of observations with the most common target variable values with the goal of better balancing the ratio between these observations and the ones with the interesting target values (which are less frequent). SMOTE works by combining under-sampling of the frequent classes with over-sampling of the minority class, through artificial generation of new cases of the minority class by interpolating between pairs of existing cases. 

Recently, these methods were extended~\cite{Torgo2013,TBRP14} for regression tasks as it is the case of our problem. The utility-based framework proposed by the authors for regression relies on the notion of relevance of the target variable which varies according to its domain. The relevance expresses the bias in the domain and is defined as a continuous function $\phi(Y): \delta \rightarrow [0,1]$ mapping the target variable from the domain $\delta$ into a $[0,1]$ scale of relevance. Being a domain-based function, this framework requires an input by the user concerning the relevance function but also the relevance threshold, from which we define the set of rare cases.

We have used the work of these authors to create two variants of each of our datasets. The first variant uses the SMOTEr algorithm to create a new training set by over-sampling the set of cases with extremely large number of tweets $\boldsymbol{D}_r$, and under-sampling the set of most frequent cases $\boldsymbol{D}_i$, thus balancing the resulting distribution of the target variable. The sets of rare and normal cases are defined using the relevance function previously mentioned, given a relevance threshold defined by the user. Concerning the artificial generation of new cases of the minority class, the process is not as trivial as in the original SMOTE algorithm where the rare cases have the same target variable. In this case, although in a pair of cases both have high relevance, they may not have the exact same target variable. To tackle this issue, the authors propose a weighted average of the target variable defining the weight as an inverse function of the distance between the generated case and the cases used.

The second variant uses the under-sampling algorithm proposed by the same authors to decrease the number of cases with low number of tweets $\boldsymbol{D}_i$, hence the most common, once again resulting in a more balanced distribution. In addition to the relevance threshold set by the user, the ratio between ''normal'' and rare cases should also be inputted by the user in order to define the amount of cases from the most common set of cases which should be randomly removed.

In our experiments we will apply and compare these methodologies in order to check which one provides better results in forecasting accurately the number of tweets of highly popular news items.

\section{Data and Methods}\label{sec:datamethods}

\subsection{The Used Data}\label{subsec:data}

\begin{figure}[!h]
\centering
\includegraphics[width=\linewidth]{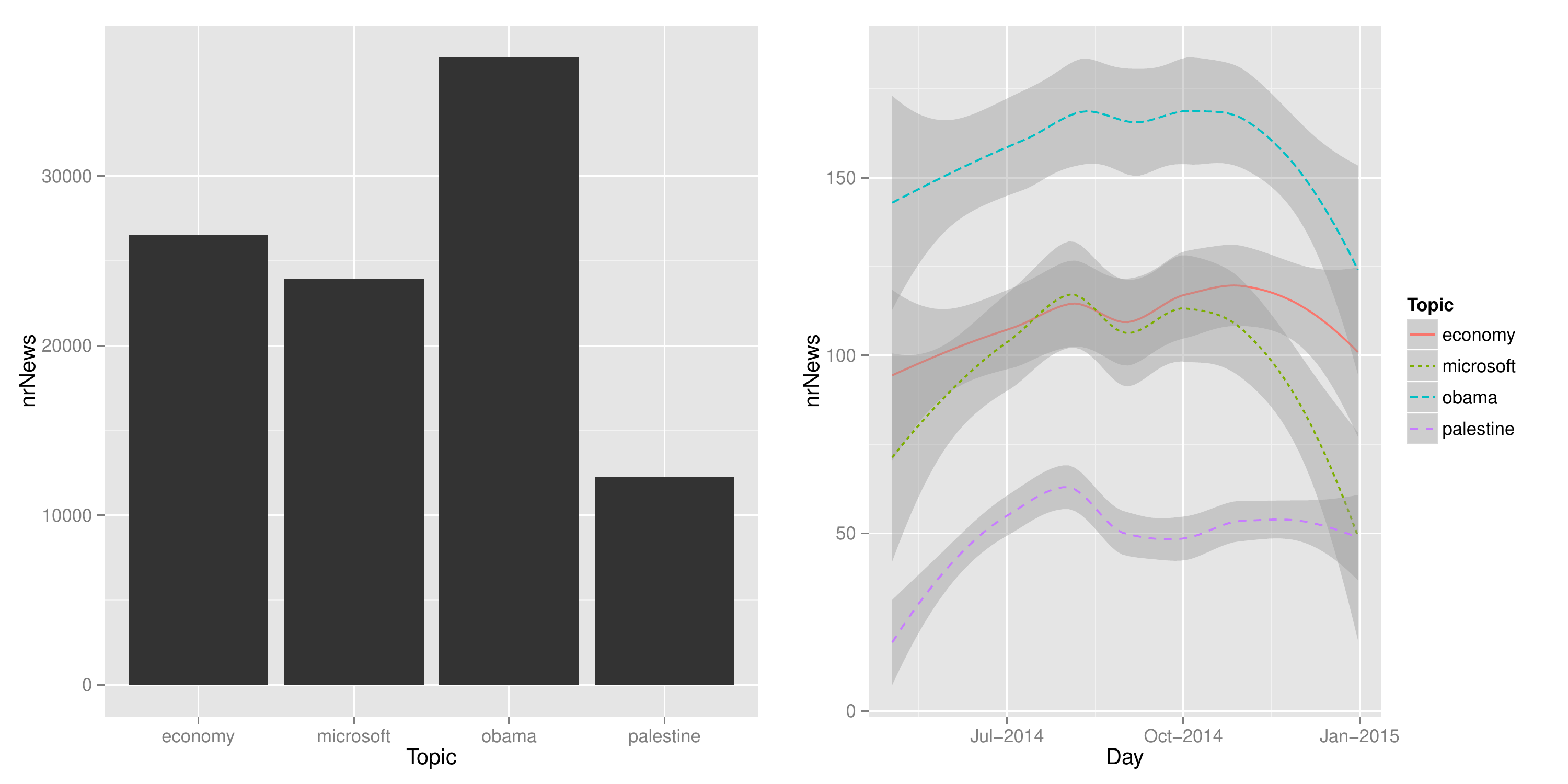}
\caption{Number of news per topic (left) and a smoothed approximation of the amount of news per day for each topic}\label{img:newstopicday}
\end{figure}

The experiments are based on news concerning four specific topics: \emph{economy, microsoft, obama,} and \emph{palestine}. These topics were chosen due to two factors: their (worldwide) popularity and the fact that they report to different types of entities (sector, company, person, and country, respectively). For each of the four topics we have constructed a dataset with news mentioned in Google News during a period spanning roughly eight months (2014-May-01 - 2015-Jan-01). The datasets were created by querying Google News every 30 minutes and collecting the top-100 news returned.\footnote{While it would be interesting to incorporate input from the news recommender system Digg, we were unable to include it in the evaluation process due to the deprecation of their API.}. Figure~\ref{img:newstopicday} shows the total number of news per topic during this period (left) and a smoothed approximation of the amount of news per day for each topic (right).\footnote{The total number of news for all topics is 106.456.}

For each news recommended by Google News the following information was collected: title, headline, publication date and its position in the ranking. For each of the four topics a dataset was built for solving the predictive task formalized in Section~\ref{subsec:dataminingtask}. These datasets were built using the following procedure. For obtaining the target variable value we have used the Twitter API\footnote{Twitter API Documentation: https://dev.twitter.com/docs/api} to check the number of times the news was tweeted in the two days following its publication. This two days' limit was decided based on the work of Yang and Leskovec~\cite{Yang2011} and our own observations, both of which suggest that after a few days the news stop being tweeted. Of the total number of news for all topics (106.456), in 6.411 cases ($6\%$) it was not possible to obtain the number of tweets and in 19.719 cases ($18,5\%$) the news items were not tweeted.

In terms of predictor variables used to describe each news we have selected the following. We have applied a standard bag-of-words approach to the news headline to obtain a set of terms describing it, using the infrastructure provided by the R package \textbf{tm} \cite{Feinerer2008}. Initial experiments have shown that the headline provides better results than the title of the news item. We have not considered the use of the full news text as this would require following the available link to the original news site and have a specific crawler to obtain this text. Given the wide diversity of news sites that are aggregated by Google News, this would be an unfeasible task. To this set of predictors we have added two sentiment scores: one for the title and the other for the headline. These two scores were obtained by applying the function \texttt{polarity()}  of the R package \textbf{qdap} \cite{Qdap2013} that is based on the sentiment dictionary described by Hu and Liu~\cite{Hu2004}.
Summarising, our four datasets are built using the information described on Table~\ref{tab:features} for each available news. 

\begin{table}
\centering
\caption{The variables used in our predictive tasks\label{tab:features}}{
\begin{tabular}{|l|p{0.6\textwidth}|}
\hline \textbf{Variable} & \textbf{Description} \\ 
\hline \textit{NrTweets}  & The number of times the news was tweeted in the two days following its publication. This is the target variable $Y$.  \\ 
\hline $T_1, T_2, \cdots$ & The term frequency of the terms selected through the bag of words approach when applied to all news headlines. \\ 
\hline \textit{SentTitle} & The sentiment score of the news title. \\ 
\hline \textit{SentHeadline} & The sentiment score of the news headline. \\ 
\hline 
\end{tabular}}
\end{table}

As expected, the distribution of the values of the target variable for the four obtained datasets is highly skewed, in accordance with previous work~\cite{Tatar2014}. Moreover, as we have mentioned our goal is the accuracy at the low frequency cases where the number of tweets is very high. We will apply the different resampling strategies described in Section~\ref{subsec:resample} to our collected data. This will lead to 12 different datasets, three for each of the selected topics: (i) the original imbalanced dataset; (ii) the dataset balanced using SMOTEr; and (iii) the dataset balanced using under-sampling. The hypothesis driving the current paper is that by using the re-sampled variants of the four original datasets we will gain predictive accuracy at the highly tweeted news, which are the most relevant for providing accurate news recommendations.

Since a news item may appear in more than one position in the Google News ranking, on different timestamps, additional datasets are built for each topic in order to harness that information. For each topic a dataset is constructed containing a news item identifier, the timestamp of the query and the respective position. This data is used in the second data mining task previously described in Section~\ref{subsec:dataminingtask}.

%================================================================
\subsection{Regression Algorithms}\label{sec:propmeth}

In order to test our hypothesis that using resampling methods will improve the predictive accuracy of the models on the cases that matter to our application, we have selected a diverse set of regression tools. Our goal here is to try to make sure our conclusions are not biased by the choice of a particular regression tool.

Table~\ref{table:regralg} shows the regression methods and tools that were used in our experiments. To make sure our work can be easily replicable we have used the implementations of these tools available in the free and open source R environment. All tools were applied using their default parameter values.

\begin{table}
\centering
\caption{Regression algorithms and respective R packages\label{table:regralg}}{
%\scalebox{0.9} {
\begin{tabular}{lll}
\hline
ID & Method & R package\\
\hline
RF & Random forests & randomForest \cite{Liaw2002}\\
LM & Multiple linear regression & stats \cite{R2014}\\
SVM & Support vector machines & e1071 \cite{Meyer2012}\\
MARS & Multivariate adaptive regression splines & earth \cite{Milborrow2013}\\
\hline
\end{tabular}}
%}
\end{table}

%================================================================
\subsection{Evaluation Metrics}\label{subsec:evalmetrics}

The metrics presented here are used to evaluate the performance of our approach in both their parts: (1) the prediction of rare cases of highly tweeted news and (2) the production of news rankings using those predictions.

\subsubsection{Prediction Evaluation Metrics}\label{subsubsec:predmodelevalmetrics}

It is a well-known fact that when the interest of the user is a small proportion of rare events, the use of standard predictive performance metrics will lead to biased conclusions. In effect, standard prediction metrics focus on the "average" behaviour of the prediction models and for these tasks the user goal is a small and rare proportion of the cases. Most of the previous studies on this type of problems was carried out for classification tasks, however, Torgo and Ribeiro \cite{Torgo2007} and Ribeiro \cite{Ribeiro2011} have shown that the same problems arise on regression tasks when using standard metrics like for instance the Mean Squared Error. Moreover, these authors have shown that discretizing the target numeric variable into a nominal variable followed by the application of classification algorithms is also prone to problems and leads to sub-optimal results.

In this context, we will base our evaluation on the utility-based regression framework proposed in the work by Torgo and Ribeiro~\cite{Torgo2007} and Ribeiro~\cite{Ribeiro2011}. The metrics proposed by these authors assume that the user is able to specify what is the sub-range of the target variable values that is most relevant. This is done by specifying a relevance function $\phi$ that maps the values of the target variable into a $[0,1]$ scale of relevance. Using this mapping and a user-provided relevance threshold the authors defined a series of metrics that focus the evaluation of models on the cases that matter for the user. In our experiments we have used as relevance threshold the value of $0.9$, which leads to having on average  7\% to 10\% of the cases tagged as rare depending on the topic.

In our prediction models' evaluation process we will mainly rely on one utility-based regression metric: F-Score. This is a composite measure that integrates the values of precision and recall according to their adaptation for regression described in the above mentioned utility-based evaluation framework. The F-Score measure is able to consider situations where the models forecast a high number of tweets for a news that ends up having a low  number of tweets, \textit{i.e.} false positives.

\subsubsection{Ranking Evaluation Metrics}\label{subsubsec:rankevalmetrics}

Considering that, as referred before, the order of presentation is relevant to our endeavours, the metrics used to evaluate the production of rankings should be able to address that issue. In this context,  we propose the use of several metrics including Precision at \textit{k} (P@\textit{k}), Average Precision (\textit{AP}), Mean Average Precision (\textit{MAP}), R-Precision (\textit{RP}), Mean R-Precision (\textit{MRP}), Mean Reciprocal Rank (\textit{MRR}) and Normalized Discounted Cumulative Gain (NDCG@\textit{k}).

These metrics may be divided into two sets: (1) those focused on the global outcome, and (2) those which take into account the position of a given item in the ranking. The first set is formed by P@\textit{k}, \textit{AP}, \textit{MAP}, \textit{RP} and \textit{MRP}. The metrics \textit{MRR} and NDCG@\textit{k} form the second set.

Precision at \textit{k} measures the number of relevant items on the top-\textit{k} ranking positions.

\begin{equation}
P@k(q) = \frac{\textit{\# of relevant items up to rank k in query q}}{k}
\end{equation}

The metric Average Precision computes the average precision for all values of \textit{k} where \textit{k} is the rank, \textit{n} is the number of retrieved items and $\textit{Rel}_k$ is a binary function evaluating the relevance of the \textit{k}th ranked item, attributing 1 to the relevant items at rank \textit{k} and 0 otherwise. The Mean Average Precision is computed to determine the effectiveness of the ranking mechanism over all queries, where $\left | Q \right |$ is the number of queries.

\begin{equation}
AP(q)=\frac{\sum _{k=1}^{n} P@k(q) \times Rel_k}{\textit{\# of relevant items for query q}}
\end{equation}

\begin{equation}
MAP=\frac{\sum _{q=1}^{\left | Q \right |} AP(q)}{\left | Q \right |}
\end{equation}

In order to tackle the MAP issue concerning the effect of equally weighting each AP value, with disregard for the number of relevant documents found in each of the queries, the measure R-Precision \cite{Croft2009} is introduced. It measures the fraction of relevant items for the query \textit{q} that are successfully retrieved at the \textit{R}th position in the ranking. The Mean R-Precision corresponds to the arithmetic mean of all the R-Precision values for the set of all queries.

\begin{equation}
RP(q) = \frac{\textit{\# Rel}_{R}}{\textit{\# of relevant items for query q}}
\end{equation}

\begin{equation}
MRP(q) = \frac{\sum _{q=1}^{\left | Q \right |} RP(q)}{\left | Q \right |}
\end{equation}

Taking into account the ranking position, the Reciprocal Rank is defined as the inverse of the rank at which the first relevant document is retrieved. As in MAP and MRP, the Mean Reciprocal Rank is defined as the average of the reciprocal ranks over all queries where $\left | Q \right |$ is its total and $rank_{q}$ is the rank position where the first relevant item\footnote{According to our objectives, we have established that the relevant items are those which belong to the top 10.} for the query \textit{q} was found.

\begin{equation}
MRR = \frac{1}{\left | Q \right |} \sum _{q=1}^{\left | Q \right |} \frac{1}{rank _{q}}
\end{equation}

Finally, Normalized Discounted Cumulative Gain measures the search result quality of the ranking function by assigning high weights to documents in highly ranked positions and reducing the ones found in lower ranks. Its definition is presented as follows, where $Rel _{i,q} \epsilon \left \{ 0,1,2,3 \right \}$ is the relevance judgment of the \textit{i}th ranked item for query q. The normalization of Discounted Cumulative Gain (DCG) to a value between 0 and 1 is done by dividing the DCG value for the ideal ordering of DCG ($idealDCG$).

\begin{equation}
DCG@\textit{k}(q) = \sum_{i=1}^{k} \frac{2 ^{Rel _{i,q}} - 1}{log _2 (1+i)}
\end{equation}

\begin{equation}
NDCG@\textit{k} = \frac{\sum _{q=1}^{Q} \frac{DCG@\textit{k}(q)}{idealDCG@\textit{k}(q)}}{Q}
\end{equation}

Concerning the concrete evaluation process we decided to use the metrics MAP, MRP, MRR and NDCG to enable a dual evaluation: a first that evaluates the global outcome and a second that takes into consideration the specific rank of each item.

\section{Experimental Evaluation}\label{sec:expeval}

This section presents the results on three sets of experiments. The first set evaluates the ability of models accurately predicting the news items with a high number of tweets. The second set of results relates to the application of proposed framework as a stand-alone system creating news rankings with the predicted number of tweets of recent news, and its evaluation. Finally, the third set is a real-world usage evaluation where our framework is applied to the news recommendations of Google News. Also, in this third set we test the possibility of augmenting the recommendations of Google News with the results of our framework to determine the ability of improving the quickness in recommending highly popular news. In all of these experiments the baseline is derived from the number of tweets assuming that our objective is high accuracy in recommending those news which have the most number of tweets.

\subsection{Prediction Models Evaluation}\label{subsec:predmodeval}

Our data (news items) has a temporal order. In this context, one needs to be careful in terms of the process used to obtain reliable estimates of the selected evaluation metrics. This means that the experimental methodology should make sure that the original order of the news is kept so that models are trained on past data and tested on future data to avoid over-optimistic estimates of their scores. In this context, we have used Monte Carlo simulation as the experimental methodology to obtain reliable estimates of the selected evaluation metrics for each of the alternative approaches. This methodology randomly selects a set of points in time within the available data, and then for each of these points selects a certain past window as training data and a subsequent window as test data, with the overall training+test process repeated for each point. All alternative approaches are compared using the same training and test sets to ensure fair pairwise comparisons of the obtained estimates. Our results are obtained through 50 repetitions of a Monte Carlo estimation process with 50\% of the cases used as training set and the subsequent 25\% used as test set. This process is carried out in R using the infrastructure provided by the R package \textbf{performanceEstimation} \cite{Torgo2013b}.

Our results cover four topics, as referred before: \emph{economy, microsoft, obama,} and \emph{palestine}. Table~\ref{table:pm-results_f1} presents a summary of the estimated metric scores for the different setups that were considered. These three metrics (precision, recall and F1) are the most interesting from the perspective of our application with emphasis in the F1 measure because it penalises false positives (\textit{i.e.} predicting a very high number of tweets for a news item that is not highly tweeted). For each regression algorithm the best estimated scores are denoted in italics, whilst the best overall score is in bold. 

\begin{table}[!h]
\centering
\caption{Precision, Recall and F1-Score estimated scores for all topics.\label{table:pm-results_f1}}{
\scalebox{0.9} {
\begin{tabular}{l|rrr|rrr|rrr|rrr}
  \cline{2-13}
\multicolumn{1}{c|}{} & \multicolumn{3}{c|}{economy} & \multicolumn{3}{c|}{microsoft} & \multicolumn{3}{c|}{obama} & \multicolumn{3}{c}{palestine} \\
& prec & rec & F1 & prec & rec & F1 & prec & rec & F1 & prec & rec & F1 \\ 
\hline
lm            & 0.17 & 0.04 & 0.07 & 0.15 & 0.03 & 0.04 & 0.11 & 0.00 & 0.01 & 0.32 & 0.10 & 0.15 \\ 
lm\_SMOTE     & \textit{0.65} & \textit{0.30} & \textit{0.41} & 0.58 & \textit{0.33} & \textit{0.42} & 0.57 & \textit{0.48} & \textit{0.52} & 0.60 & \textit{0.20} & \textit{0.29} \\ 
lm\_UNDER     & \textit{0.65} & 0.28 & 0.40 & \textit{0.60} & 0.32 & 0.41 & \textit{0.58} & \textit{0.48} & \textit{0.52} & \textit{0.62} & 0.18 & 0.28 \\ 
\hline
svm           & 0.10 & 0.00 & 0.00 & 0.03 & 0.00 & 0.00 & 0.07 & 0.00 & 0.00 & 0.03 & 0.00 & 0.01 \\ 
svm\_SMOTE    & 0.61 & 0.44 & 0.51 & \textit{0.64} & 0.56 & \textit{0.60} & 0.52 & 0.55 & 0.53 & \textbf{0.74} & \textbf{0.49} & \textbf{0.59} \\ 
svm\_UNDER    & \textit{0.68} & \textbf{0.47} & \textit{0.56} & \textit{0.64} & \textbf{0.57} & \textit{0.60} & \textit{0.58} & \textbf{0.59} & \textit{0.58} & 0.69 & \textbf{0.49} & 0.57 \\ 
\hline
mars          & 0.15 & 0.02 & 0.04 & 0.14 & 0.01 & 0.02 & 0.11 & 0.00 & 0.00 & 0.38 & 0.05 & 0.09 \\ 
mars\_SMOTE   & 0.68 & 0.40 & 0.50 & 0.61 & 0.44 & 0.51 & 0.58 & 0.54 & 0.55 & 0.61 & 0.29 & 0.39 \\ 
mars\_UNDER   & \textbf{0.77} & \textbf{0.47} & \textbf{0.58} & \textbf{0.74} & \textit{0.52} & \textbf{0.61} & \textbf{0.64} & \textbf{0.59} & \textbf{0.61} & \textit{0.71} & \textit{0.37} & \textit{0.48} \\ 
\hline
rf            & 0.27 & 0.05 & 0.08 & 0.25 & 0.04 & 0.06 & 0.20 & 0.02 & 0.04 & 0.27 & 0.06 & 0.10 \\ 
rf\_SMOTE     & 0.64 & \textit{0.43} & 0.51 & 0.56 & 0.47 & 0.51 & 0.52 & 0.52 & 0.52 & 0.62 & \textit{0.36} & 0.45 \\ 
rf\_UNDER     & \textit{0.70} & 0.42 & \textit{0.52} & \textit{0.65} & \textit{0.49} & \textit{0.56} & \textit{0.60} & \textit{0.56} & \textit{0.57} & \textit{0.69} & 0.35 & \textit{0.47} \\ 
\hline
\end{tabular}}
}
\end{table}

These results show that in all setups every algorithm is able to take advantage of resampling strategies to boost its performance. The results obtained with Random Forest, MARS and SVMs are encouraging, moreover taking into account that all methods were applied with their default parameter settings. With F1 scores around $0.6$ we can have more confidence that if we use the predictions of these models for ranking news items by their predicted number of tweets, the resulting rank will match reasonably well the preferences of the users.

Overall, the main conclusion from our comparisons is that resampling methods are very effective in improving the predictive accuracy of different models for the specific task of forecasting the number of tweets of highly popular news. These methods are able to overcome the difficulty of these news being infrequent. This is particularly important within our goal, which requires the ability to accurately identify the news that are more relevant for the users in order to improve the performance of news recommender systems.

\subsection{Rankings Evaluation}\label{subsec:rankeval}

The goal of the evaluation presented in this section is to check how effective are the news recommendation rankings produced using the proposed framework. Therefore, we will want to compare the rankings produced by our method against the ground truth. The ground truth rank is based on the observed number of tweets of each news item within the period of two days following its publication timestamp. Based on these comparisons we will calculate three of the metrics described in Section~\ref{subsubsec:rankevalmetrics}, Mean Average Precision ($MAP$), Mean Reciprocal Rank ($MRR$) and Normalized Discounted Cumulative Gain ($NDCG@k$). With that goal in mind we have designed the following experiment to compare the two news recommendations.

Concerning the prediction models and their train and test sets for this evaluation, we obtained our results through 100 repetitions of a Monte Carlo estimation process with 20\% of the cases used as training set ($Tr$) and the following 24 cases\footnote{Corresponds to a time span of half-day.} used as test set\footnote{In effect, we remove from the training set the news whose publication date is greater than the end of the training period less 2 days. These news are too recent and thus we still do not know their number of tweets (we count it after 2 days) and thus are useless for training our models.} ($Ts$).

Each time window ($Tr+Ts$) used in a Monte Carlo iteration is split in 30 minutes time intervals and we obtain the top 100 news recommendations at each of these steps from Google News. This leads to a data set as shown in Table~\ref{table:modifiedrankings}, where, contrary to the previous evaluation where the cases were the news items, the rows (top 100 news items recommended by Google News) represent the cases referred. Each time $Q_i \in Ts$ forms our test set where we want to compare our proposed ranking against the ground truth. This comparison is carried out as follows. 

\begin{table}
\centering
\caption{Illustration of the dataset used in rankings evaluation\label{table:modifiedrankings}}{
\begin{tabular}{|l|l|l|l|l|}
\hline
\textbf{Timestamp} & \textbf{$R_1$} & \textbf{$R_2$} & \textbf{...} & \textbf{$R_{100}$}\\ 
\hline 
$Q_1$ & $N_{14}$ & $N_{43}$ & ... & $N_{73}$ \\ 
\hline
$Q_2$ & $N_{43}$ & $N_{14}$ & ... & $N_{85}$ \\ 
\hline 
$Q_3$ & $N_{43}$ & $N_{23}$ & ... & $N_{85}$ \\ 
\hline 
$Q_t$ & ... & ... & ... & ... \\ 
\hline 
\end{tabular}}
\end{table}

The starting point of our method is a news pool ($NP_i$) of all the news items ranked by Google News belonging to the cases in the test set, $Ts$. This set can be decomposed into the set of news with publication date in the training period\footnote{For these we already know their observed number of tweets as two days have already passed since their publication date.} ($NP_i^{old}$) and the set of news for which we still do not know the number of tweets ($NP_i^{new}$) because two days have not yet passed since their publication date. To our endeavour in this evaluation we will only use the news items of $NP_i^{new}$ as our objective is to ascertain the ability of our prediction models to produce correct rankings.

Using the observed number of tweets of each of the items in $NP_i^{new}$, a ground truth ranking is obtained ($TR_i$). The predicted ranking ($PR_i$) is built using our models to predict and rank the number of tweets of each news item in $NP_i^{new}$. After obtaining both the ground truth ($TR_i$) and the predicted ($PR_i$) rankings, the ranking evaluation metrics are calculated and the results are presented in Table~\ref{table:rank-results}.

\begin{table}
\centering
\caption{MAP, MRR and NDCG@10 scores for all topics\label{table:rank-results}}{
%\scalebox{0.9} {
\begin{tabular}{l|ccc|ccc}
  \cline{2-7}
\multicolumn{1}{c|}{} & \multicolumn{3}{c|}{economy} & \multicolumn{3}{c}{microsoft}\\
  \hline
 & MAP & RR & NDCG10 & MAP & RR & NDCG10 \\ 
  \hline
svmSMOTE & 0.42 & 0.59 & 0.56 & 0.56 & 0.67 & 0.65 \\ 
svmUNDER & \textbf{0.44} & \textbf{0.65} & \textbf{0.57} & 0.57 & 0.72 & 0.66 \\ 
marsSMOTE & 0.42 & 0.62 & 0.55 & 0.57 & \textbf{0.74} & 0.65 \\ 
marsUNDER & 0.40 & 0.58 & 0.55 & 0.57 & 0.71 & 0.66 \\ 
rfSMOTE & 0.43 & \textbf{0.65} & \textbf{0.57} & \textbf{0.58} & \textbf{0.74} & \textbf{0.68} \\ 
rfUNDER & 0.42 & 0.62 & 0.56 & \textbf{0.58} & 0.73 & 0.67 \\ 
   \hline
\multicolumn{1}{c|}{} & \multicolumn{3}{c|}{obama} & \multicolumn{3}{c}{palestine} \\
    \hline
svmSMOTE & 0.26 & 0.47 & 0.40 & 0.80 & 0.89 & 0.86 \\ 
svmUNDER & \textbf{0.28} & \textbf{0.48} & 0.40 & 0.80 & 0.90 & 0.85 \\ 
marsSMOTE & 0.26 & 0.45 & 0.39 & 0.81 & \textbf{0.93} & 0.86 \\ 
marsUNDER & 0.27 & 0.46 & 0.39 & 0.79 & 0.89 & 0.85 \\ 
rfSMOTE & 0.27 & \textbf{0.48} & \textbf{0.41} & \textbf{0.82} & \textbf{0.93} & \textbf{0.87} \\ 
rfUNDER & 0.26 & 0.47 & 0.40 & 0.79 & 0.88 & 0.85 \\ 
   \hline
\end{tabular}}
%}
\end{table}

Results show that our approach is capable of producing news rankings which obtain good results in at least three of the four topics evaluated. Additionally, results also show that the regression algorithm Random Forest in combination with the resampling strategy SMOTE presents the best overall results, contrary to the results in the previous evaluation of the prediction models where the combination of the regression algorithm MARS with the under-sampling strategy achieved the best overall results. Furthermore, the results obtained show that it is possible to accurately rank sets of recent news items, thus reducing the latency period related to their recency for a considerable amount of cases.

\subsection{Real World Usage Comparison}\label{subsec:realworldcomp}

In this section we present an evaluation of a real world case scenario where our predicted rank ($PR_i$) is derived from the news items proposed by Google News at a given time ($GN_i$) against the ground truth. Our evaluation is focused on the top 10 news items as our objective is to check the ability to predict the users' reading preference on highly tweeted news items. Additionally, we evaluate the enhanced Google News rankings, augmented by the combination of these with the results of our predicted ranks.

The design of these experiments experiment is very similar to the previous one. However, since our news' pool is based on the news recommended by Google News at a given time, some alterations were made. They are described as follows.

\subsubsection{Google News Rankings}
For this experiment, the starting point of our method is the top 100 news obtained from Google News for time $Q_i$, $GN_i$. This set can be decomposed into $GN_i^{old}$, the news with a time-alive greater than two days, and $GN_i^{new}$, the news with a time-alive lesser than two days, as in the previous experiment. We build our predicted rank ($PR_i$) by obtaining the number of tweets for all news in $GN_i$. For those belonging to $GN_i^{old}$ we use the known number of tweets, whilst for those in $GN_i^{new}$ we use our models to predict this number of tweets. The predicted rank $PR_i$ is then compared against the ground truth that is obtained using the observed number of tweets of all news in time step $i$. A concrete illustration of this process is shown in Table~\ref{table:framework-example} where news items belonging to $GN_i^{old}$ are shown in italic.

\begin{table}
\centering
\caption{Prepared evaluation data example (top 10 of 100) for a given 30 minutes query.\label{table:framework-example}}{
\begin{tabular}{rrrr}
  \hline
GroundTruthRank & RealNTweets & PR.rankPosition & Pred.NTweets \\ 
  \hline
\textbf{1} & 474 & 49 & 127.45 \\ 
\textit{\textbf{2}} & \textit{398} & \textbf{\textit{1}} & \textit{398} \\ 
\textit{\textbf{3}} & \textit{299} & \textbf{\textit{2}} & \textit{299} \\ 
\textbf{4} & 283 & 38 & 136.21 \\ 
\textbf{5} & 278 & 52 & 125.52 \\ 
\textbf{6} & 271 & \textbf{3} & 298.59 \\ 
\textbf{7} & 270 & \textbf{9} & 244.09 \\ 
\textbf{8} & 245 & 74 & 102.24 \\ 
\textbf{9} & 198 & 29 & 157.89 \\ 
\textbf{10} & 179 & 67 & 115.33 \\ 
   \hline
\end{tabular}}
\end{table}

These comparisons are carried out for all time steps $i \in Ts$ and in the end we calculate the ranking evaluation metrics. This process is repeated 50 times with Monte Carlo estimations having 20\% of the cases used as training set ($Tr$) and the following 96 cases\footnote{Corresponds to a time span of two days.} used as test set ($Ts$). The resulting metric scores are presented in Table~\ref{table:realworld-results}.

\begin{table}
\centering
\caption{MAP, MRP, MRR and NDCG@10 scores for all topics\label{table:realworld-results}}{
%\scalebox{1} {
\begin{tabular}{l|cccc|cccc}
  \cline{2-9}
\multicolumn{1}{c|}{} & \multicolumn{4}{c|}{economy} & \multicolumn{4}{c}{microsoft}\\
  \hline
& MAP & MRP & MRR & NDCG10 & MAP & MRP & MRR & NDCG10 \\ 
  \hline
svmSMOTE & \textbf{0.74} & \textbf{0.65} & 0.84 & 0.72 & \textbf{0.76} & 0.65 & \textbf{0.90} & 0.73 \\ 
svmUNDER & \textbf{0.74} & 0.64 & \textbf{0.85} & \textbf{0.73} & \textbf{0.76} & \textbf{0.67} & 0.88 & \textbf{0.74} \\ 
marsSMOTE & 0.58 & 0.41 & 0.83 & 0.58 & 0.59 & 0.42 & 0.81 & 0.59 \\ 
marsUNDER & 0.59 & 0.43 & 0.82 & 0.59 & 0.61 & 0.46 & 0.82 & 0.61 \\ 
marsSMOTE & 0.58 & 0.41 & 0.83 & 0.58 & 0.59 & 0.42 & 0.81 & 0.59 \\ 
marsUNDER & 0.59 & 0.43 & 0.82 & 0.59 & 0.61 & 0.46 & 0.82 & 0.61 \\ 
   \hline
\multicolumn{1}{c|}{} & \multicolumn{4}{c|}{obama} & \multicolumn{4}{c}{palestine} \\
    \hline
svmSMOTE & 0.36 & 0.30 & \textbf{0.61} & 0.38 & 0.75 & 0.66 & \textbf{0.86} & 0.75 \\ 
svmUNDER & \textbf{0.37} & \textbf{0.31} & \textbf{0.61} & \textbf{0.40} & \textbf{0.76} & \textbf{0.68} & \textbf{0.86} & \textbf{0.76} \\ 
marsSMOTE & 0.31 & 0.23 & 0.56 & 0.35 & 0.66 & 0.45 & \textbf{0.86} & 0.66 \\ 
marsUNDER & 0.33 & 0.27 & 0.57 & 0.36 & 0.66 & 0.50 & 0.84 & 0.66 \\ 
marsSMOTE & 0.31 & 0.23 & 0.56 & 0.35 & 0.66 & 0.45 & \textbf{0.86} & 0.66 \\ 
marsUNDER & 0.33 & 0.27 & 0.57 & 0.36 & 0.66 & 0.50 & 0.84 & 0.66 \\ 
   \hline
\end{tabular}}
%}
\end{table}

Results show that our approach, when applied to Google News recommendations, is capable of producing news rankings using the prediction of the future importance of news items, with fairly good results in at least three of the four topics. Moreover, it should be noted that although the regression algorithm MARS in combination with the resampling strategy under-sampling obtained the best results in the prediction models evaluation, it did not produce the most accurate rankings. Also, when comparing with the results from the evaluation of our proposed framework in a stand-alone deployment, the best results where obtained with the combination of regression algorithm Random Forest in combination with the resampling strategy SMOTE. Instead, we observe that for the scenario presented in this evaluation, the combination of regression algorithm SVM and the under-sampling strategy did in fact produce the best overall results. 

\subsubsection{Google News Rankings Augmentation}

Given the previous results, we propose to evaluate the ability of combining the rankings of Google News with those provided by our framework in order to ascertain the ability of augmenting the rankings of the former with social information. This evaluation is focused on the ability of quickly recommending news which will obtain a high number of tweets.

As previously stated, the algorithm used by Google News is not public, although there is public information which allows us to have a general outline of its process. In order to carry out these experiments some alterations must be made in order to provide a fair comparison and evaluation. As stated in the patent of the Google News algorithm, a decay factor associated with the time-alive of the news items is taken into consideration. Since we do not have the necessary information to accurately emulate Google News, we introduce a linear decay factor in our framework. Given the timespan of two days previously stated we divide it into periods of 30 minutes named time slices, $t$. Therefore, each news item for which we were able to obtain information contains 96 timeslices (\textit{i.e.} two days). The decay factor $d$ for a news item at a given time slice $t$ is then formulated as

\begin{equation}
d_{t} = 1 - \frac{t-1}{t_f},
\end{equation}

where $t_f$ represents the final timeslice ($t=96$). This factor is applied both to the ground truth and to the results of our framework along the evaluation process.

We propose two alternatives regarding the combination of both our framework and Google News' rankings: \textit{i)} by calculating the average rank of each respective news item in both ranks (\textit{agreement}), and \textit{ii)} by choosing the minimum rank for each respective news items in both ranks (\textit{poll}). For the purposes of this evaluation we use the best ranking model according to our previous evaluation, which is the combination of the SVM regression algorithm with the under-sampling strategy. Additionally, we use a baseline corresponding to the rank proposed by Google News ordered by time-alive, where the most recent news are placed on the top of the ranking.

\begin{figure}[!h]
\centering
\includegraphics[width=\linewidth]{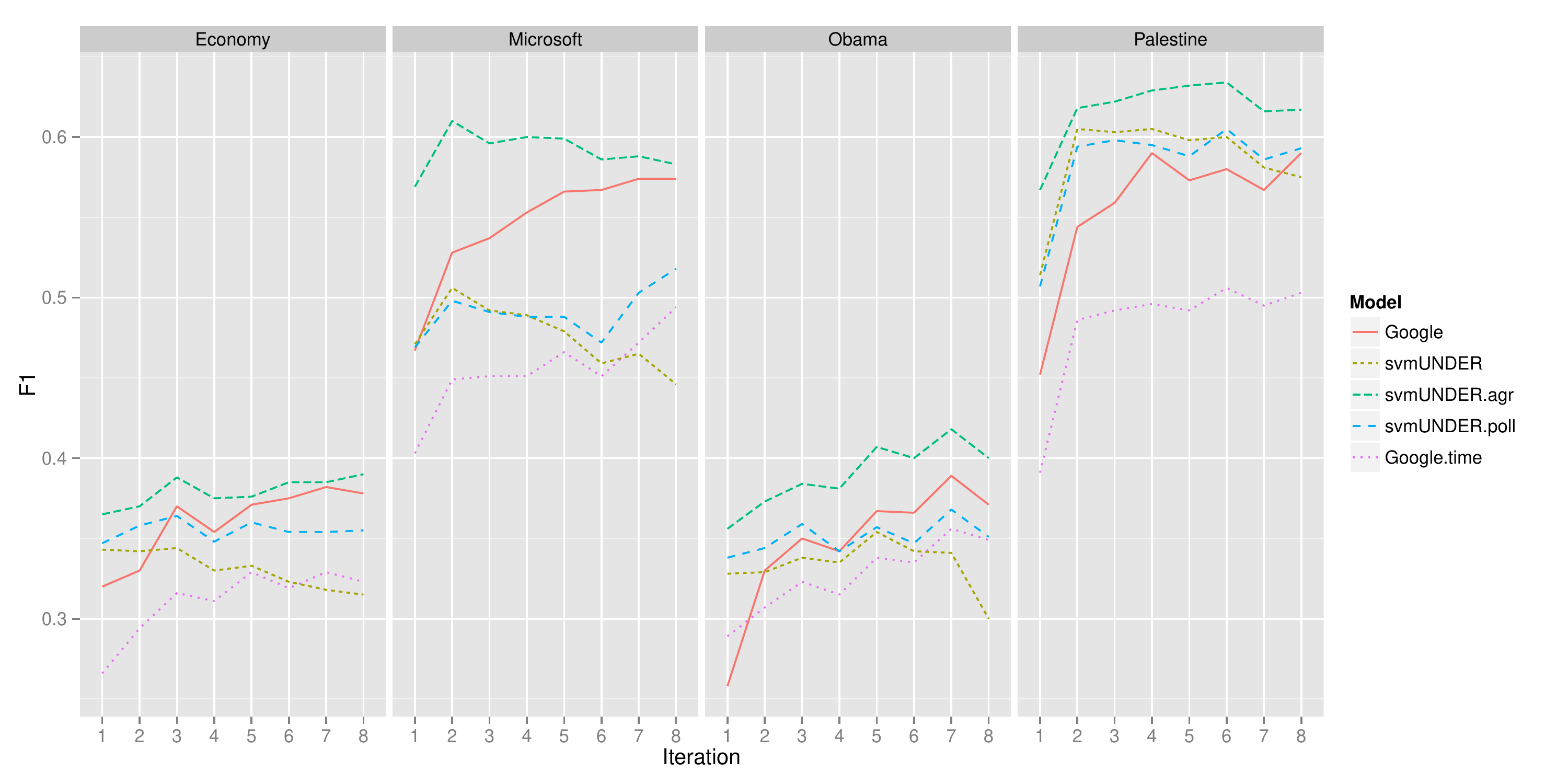}
\caption{F1 scores of the proposed approaches for all highly tweeted news in all topics in the first four hours after being published.}\label{img:eval4}
\end{figure}

The evaluation process is carried out in the same manner as the previously described evaluation, altering our focus point to the first 4 hours of the alive-time of the news items ($t \in \left \{ 1,2,...,8 \right \}$), in order to evaluate the ability of the various approaches to accurately propose news rankings in the first moments after their publication. For the purposes of this evaluation we use the metric F1, which as previously described takes into consideration both precision and recall, and penalizes false positives (\textit{i.e.} news which are predicted to have a high number of tweets but do not). The results presented in Figure~\ref{img:eval4} report to the alternatives previously described, namely the combined rankings (\textit{svmUNDER.agr} and \textit{svmUNDER.poll}), the stand-alone rankings (\textit{Google} and \textit{svmUNDER}), and the baseline rank defined as the rank proposed by Google News ordered by time-alive (\textit{Google.time}).

The results show that the combined rankings, and more specifically the agreement strategy (\textit{svmUNDER.agr}), outperforms all other approaches in predicting the popular news items. This is true for the poll strategy (\textit{svmUNDER.poll}) but only when considering the first moments after the news items' publication. Also, we observe that simply proposing the most recent news items (\textit{Google.time}) provides the worst results. Moreover, we observe that our framework (\textit{svmUNDER}) manages to better identify important news during the first moments after its publication when comparing to recommendations of Google News, although, it should be stated that there is no data that would corroborate that this is the prime objective of the known news recommender system.

\section{Conclusions}\label{sec:conclusions}

We have presented an approach to news recommendation that aims at predicting the reading interests of users of these services. This approach uses the number of times a news is tweeted as a proxy for its relevance for users and, in this context, tries to predict this number for very recent news items. Given that these predictions will be used to rank news, the models should focus on decreasing the prediction error for highly tweeted news, which are rare cases. This fact leads to one of the main contributions of our work: the study and comparison of modelling techniques that are able to accurately forecast news items with a high number tweets of news items.
The importance of this prediction is to minimize the referred latency and enable the treatment and recommendation of recent news at any time. We propose a framework for the translation of these predictions into rankings, as well as their evaluation. This particular instantiation of the framework uses two data sources: Google News and Twitter. 
The evaluation processes of both prediction models and rankings recommended by the proposed framework demonstrated that it is possible to successfully approach and tackle the issue of latency related to the recency of news items, producing more robust solutions that are capable of taking into consideration the users' reading interests. 
Concerning the proposal of rankings, we evaluated our framework in two different scenarios of deployment: \textit{1)} as a stand-alone system containing a news pool of recently published news, and \textit{2)} being applied to the news recommendations of Google News. In both cases, successful solutions where found that obtained fairly good results. Additionally, we proposed an approach to the augmentation of official news recommendations with social information. This was carried out by combining the results of Google News with the outcome of our proposed framework. Upon the evaluation of the alternatives proposed for this combination process, we concluded that our combinatorial solutions provides advantages to both the Google News and our framework alone.
Finally, concerning future work, it is our intention to broaden the basis of analysis in order to include information from multiple official and social media sources.

\section{Acknowledgments}

This work is financed by the FCT – Fundação para a Ciência e a Tecnologia (Portuguese Foundation for Science and Technology) within project UID/EEA/50014/2013. The work of N. Moniz is supported by a PhD scholarship of FCT (SFRH/BD/90180/2012).
The authors would like to thank Ricardo Campos and Fátima Rodrigues for their comments and inputs.

\bibliographystyle{abbrv}      % basic style, author-year citations
\bibliography{bibdb}   % name your BibTeX data base

\end{document}